\documentclass[final,3p,times]{elsarticle}

\usepackage{amssymb}

\usepackage{amsmath}
\usepackage{physics}
\usepackage{algorithm}
\usepackage{algorithmic}
\usepackage{array}
\usepackage{booktabs}
\usepackage{placeins}
\usepackage[T1]{fontenc} 
\usepackage[font=small,labelfont=bf]{caption}
\usepackage{tikz}

\makeatletter
\def\ps@pprintTitle{%
  \let\@oddhead\@empty
  \let\@evenhead\@empty
  \let\@oddfoot\@empty
  \let\@evenfoot\@oddfoot
}
\makeatother

\journal{Theoretical Computer Science}

\begin{document}

\begin{frontmatter}

  \title{Quantum DNA Sequencing using Gaussian Amplitude Amplification}

 \author[add1]{Richard Marin \corref{cor1}}
\ead{ramarin@mymail.mapua.edu.ph}
\author[add1,add2]{Carlos F. Baldo III \corref{cor1}}
\ead{ciiifbaldo@mapua.edu.ph}

\address[add1]{Department of Physics-Mapua University, Muralla street, Intramuros, Manila, Philippines 1002}
\address[add2]{Asia Pacific Center for Theoretical Physics, POSTECH campus, Pohang, Gyeongsangbuk, Korea 37673}

  
  \cortext[cor1]{Corresponding author}

  \begin{abstract}
    In this study, we explore how quantum pathfinding algorithm called Gaussian Amplitude Amplification (GAA) can be used to solve the DNA sequencing problem. To do this, sequencing by hybridization was assumed wherein short fragments of the nucleic acids called oligonucleotides of length $l$ were gathered and were then assembled. The process of reassembling the sequence was then abstracted into a graph problem of finding the Hamiltonian path with the least cost. The constructed directed graph was then converted into sequential bipartite graphs in order to use GAA.  The results of our simulation revealed that for the case where $l=2$ and spectrum size of $|S|=4$, the probability of finding the optimal solution (with least cost) is approximately 70.92\% - a significant improvement compared to 4.17\% when the path is chosen randomly. While this study only focused on the ideal scenario where there are no errors in the spectrum, the outcomes presented here demonstrates the plausibility of using GAA as a genome sequencing method.
  \end{abstract}


\end{frontmatter}

\section{Introduction}
\label{sec:intro}

DNA sequencing is the process of determining the sequence of chemical bases present in a DNA strand. It was first introduced by F. Sanger et al. in 1977 \citep{sangerDNASequencing1977} and has been superseded by next-generation sequencing methods (NGS) \cite{mardisDNASequencingTechnologies2017}. Some of the well-known NGS methods include sequencing by synthesis (SBS), sequencing by ligation (SBL), and single-molecule sequencing (SMS) \citep{mardisDNASequencingTechnologies2017}. There have been several applications of DNA sequencing, including De Novo Genome Assembly, Genome Resequencing, Molecule Counting, and Metagenome Resequencing \citep{shendureDNASequencing402017}.

\subsection{DNA Sequencing method}
There are several types of DNA sequencing methods, each with its strengths and weaknesses  depending on the application, and selecting a specific method significantly affects the success of an experiment. The method that shall be utilized in this study is Sequencing by Hybridization (SbH). It is a DNA sequencing method in which the oligonucleotides are \textit{hybridized} under the conditions that the detection of the complementary sequences in the target nucleic acid is permitted \citep{blazewiczDNASequencingPositive1999, jamesDeoxyribonucleic2016}. SbH can be described as having two stages: a hybridization experiment where short fragments of the nucleic acids called oligonucleotides of length $l$ are gathered, and a computational step where the DNA is reassembled using the oligonucleotides collected from the hybridization experiment. There will be a total of $4^l$ possible oligonucleotides (to be called $l$-mer), since there are four chemical bases and a chain is $l$-long. The set of $l$-mers obtained from an experiment is called the spectrum. In an ideal experiment, the size of the spectrum is given by the expression, 
\begin{equation}
	\label{eq:spec_len}
	|S|=n-l+1,
\end{equation}

\noindent where $n$ is the length of a particular DNA sequence $N$. Consider, for example,  the sequence $N=\text{ACGTG}$; in an ideal experiment where $l=2$, the sequence will produce the spectrum, $S = \{\text{AC},\text{CG},\text{GT},\text{TG}\}$, with a size of $|S|=5-2+1=4$.

The cases where an experiment yields less than or more than $n-l+1$ $l$-mers are regarded as non-ideal cases with negative and positive errors, respectively. Negative errors may occur due to experimental problems or the structure of $N$. For instance, consider the sequence $N=\text{ACCCG}$. In an ideal experiment where $l=2$, the sequence can only produce the spectrum $S=\{\text{AC},\text{CC},\text{CG}\}$ despite the original sequence having $n=5$. On the other hand, positive errors only occur due to experimental problems. To show this, we take the sequence $N=\text{ACGTG}$. This experiment yields the spectrum $S=\{\text{AC},\text{GC},\text{CG},\text{GT},\text{TG}\}$, with a positive error.

\subsection{Sequencing by Hybridization}

SbH is a two-stage process involving a physical and a computational stage. The physical stage involves obtaining short polymer sequences of known length $l$ ($l$-mer) from a sequence $N$ of length $n$ to form a \textit{spectrum}. In an ideal SbH experiment, the number of $l$-mers in a spectrum is equal to $|S|$ as shown in Eq.~(\ref{eq:spec_len}). The $l$-mers in the spectrum are then analyzed in the computational stage for reconstructing the target DNA sequence \citep{mirzabekovDNASequencingHybridization1994}.

\begin{figure}[h]
	\centering
	\includegraphics[width = 14cm]{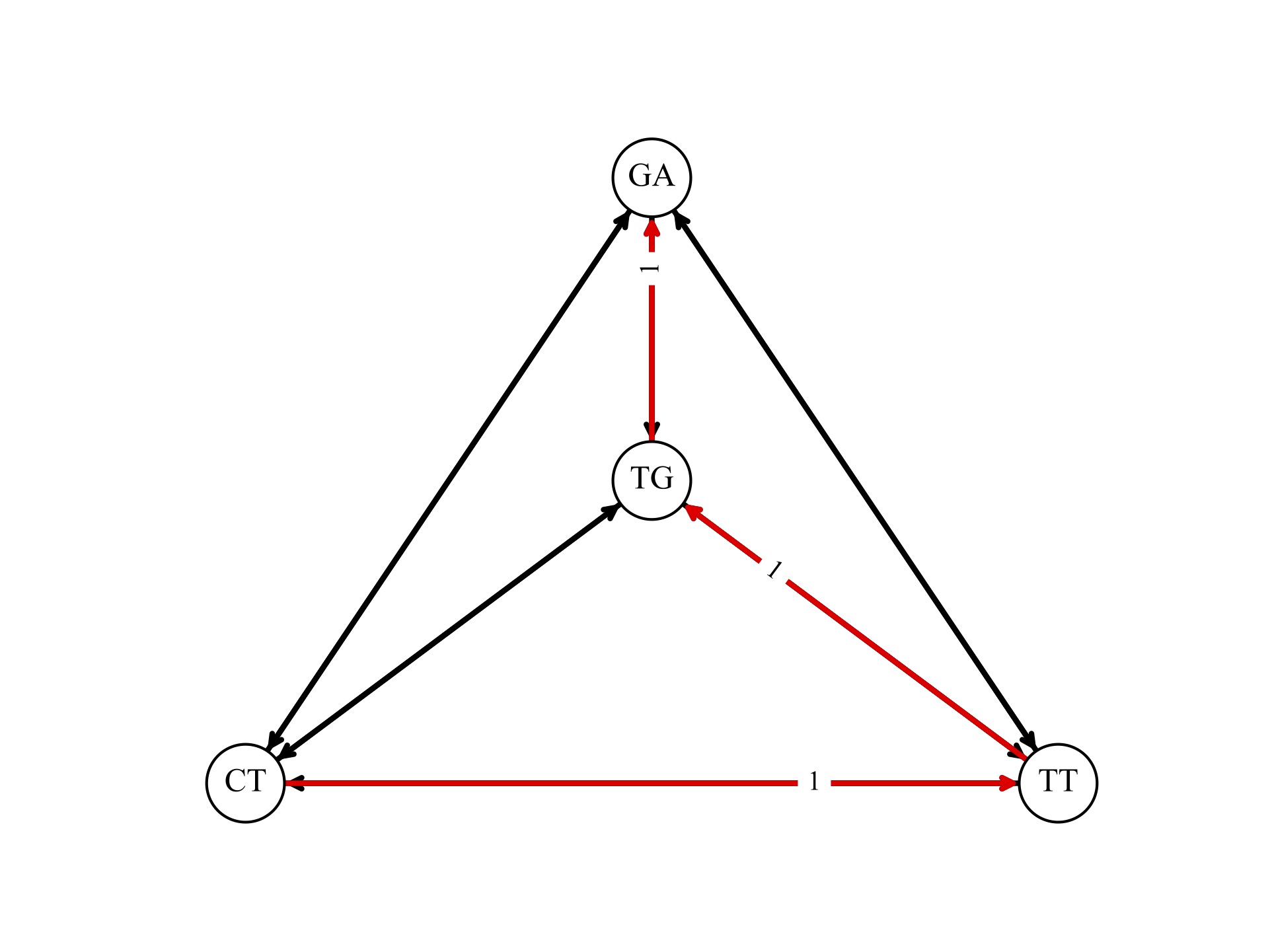}
	\caption{A Graph with the nodes being an element of the spectrum $S=\{\text{CT},\text{TT},\text{TG},\text{GA}\}$. The path with the least cost and would also reconstruct the DNA sequence is highlighted in red. The total cost of the red path is 4.}
	\label{fig:planar_graph}
\end{figure}

Reassembling the DNA sequence $N$ in the computational part of the SbH can be reframed as finding the Hamiltonian path of a graph. Consider the sequence $N=\text{CTTGA}$ with the spectrum $S=\{\text{CT},\text{TT},\text{TG},\text{GA}\}$. Fig. \ref{fig:planar_graph} shows that a graph can be constructed using $S$ by finding the path with the least cost on a graph where the nodes are the elements of $S$. The cost of connecting two $l$-mers, $C$, can be defined as $l$ minus the number of overlapping bases. For example, joining together $\text{CT}$ and $\text{TT}$ will cost $C=2-1=1$. On the other hand, joining together $\text{TT}$ and $\text{CT}$ will cost $C=2-0=2$. Notice from these examples that the cost also depends on the order in which the $l$-mers are joined. In an ideal hybridization experiment, the DNA sequence can be reconstructed in polynomial time \citep{blazewiczComplexityDNASequencing2003}. However, in the non-ideal case, the problem becomes unsolvable in polynomial time or NP-complete.

Classical algorithms have been developed to assist in reconstructing the DNA for SbH’s computational stage. In an ideal case of SbH, the problem of reconstructing the sequence can be solved by reframing the problem as finding the Eulerian path in a graph \citep{pevzner1TupleDNASequencing1989, pevznerEulerianPathApproach2001}. An Eulerian path is essentially a trail in a finite graph that visits every edge exactly once. Furthermore, the Eulerian path of a graph can be found in linear time using Hierholzer’s algorithm \citep{fleischnerAlgorithmsEulerianTrails1990}. However, the problem in the non-deal case is finding the Hamiltonian path. The Hamiltonian path is a path on a finite graph that visits every node exactly once. Finding a Hamiltonian path in a graph is akin to the traveling salesman problem (TSP), which cannot be solved in polynomial time by a classical computer (NP-Hard). Thus, the time required to find the optimal path increases superpolynomially with the number of $l$-mers in the spectrum \citep{blazewiczComplexityDNASequencing2003}. Classical algorithms have been developed for non-ideal cases \citep{blazewiczDNASequencingPositive1999, casertaHybridAlgorithmDNA2014}; however, their time complexity is still superpolynomial in the worst-case scenario.

The progress in quantum computing research has shown that a class of problems is more efficient to solve in a quantum computer than in a classical computer. Some examples of algorithms that have been developed are Shor’s prime factorization algorithm, Quantum Fourier transform, and Quantum Variational Eigensolvers \citep{gyongyosiSurveyQuantumComputing2019}. Quantum computers’ hardware fundamentally differs from classical computers since they use quantum bits (qubits) instead of classical bits. In contrast to classical bits, which can only be in the state 0 or 1, qubits can be in a superposition of 0 and 1. A quantum computer's advantage over its classical counterpart is its ability to create superpositions, allowing it to perform parallel computations.

One algorithm that is most relevant to this study is Grover’s algorithm (GA). GA was first described as a quantum database search \citep{groverFastQuantumMechanical1996}. To briefly explain how GA works, it starts by preparing a superposition of a system's possible states; then, it singles out the correct answer based on some criteria specified by a user using an oracle. GA singles out the correct answer by applying a phase on the target answer and boosts the probability of that target. This study focuses on implementing a variant of GA, the Gaussian amplitude amplification (GAA) \citep{kochGaussianAmplitudeAmplification2022}. GAA works similarly to GA but has a modified oracle that allows boosting the amplitude of a range of answers instead of a single one –- this leads to the effect that allows one to use GAA as a pathfinding algorithm that could also be used to solve the TSP.

In this study, the goal is to demonstate how GAA introduced by \citet{kochGaussianAmplitudeAmplification2022} can be used to generate optimal solutions for the DNA sequencing problem. Particularly, we aim to accomplish the following: (i) generate a directed graph with associated costs from a set of $l$-mers in the spectrum, (ii) convert the generated directed graph into sequentially connected bipartite graphs, and (3) use GAA to boost the probability of measuring a path with the least cost. our work will, however, be limited to applying the method to the ideal case of SbH where $|S|=n-l+1$, which does not consider the positive and negative errors that could be introduced in the first stage of SbH. Additionally, the results will only include $l$-mers of length $l=2$. Nevertheless, this work shall demonstrate how quantum computing could be used for DNA sequencing.

\section{Conceptual Framework}
\label{sec:conceptual_framework}

Gaussian Amplitude Amplification (GAA) was introduced by \citet{kochGaussianAmplitudeAmplification2022} as a quantum pathfinding algorithm. It works with the same principle as Grover's Algorithm \citep{groverFastQuantumMechanical1996} with the difference in how the oracle is constructed. GAA is further explained in section \ref{sec:GAA}.

Several methods exist to sequence a strand of DNA. In this study, the sequencing method called Sequencing by Hybridization (SbH) was employed. SbH is different from other forms of sequencing because using SbH makes it possible to abstract the DNA Sequencing Problem into a Graph Pathfinding Problem. The specifics of SbH will be further discussed in Section \ref{sec:SbH}.

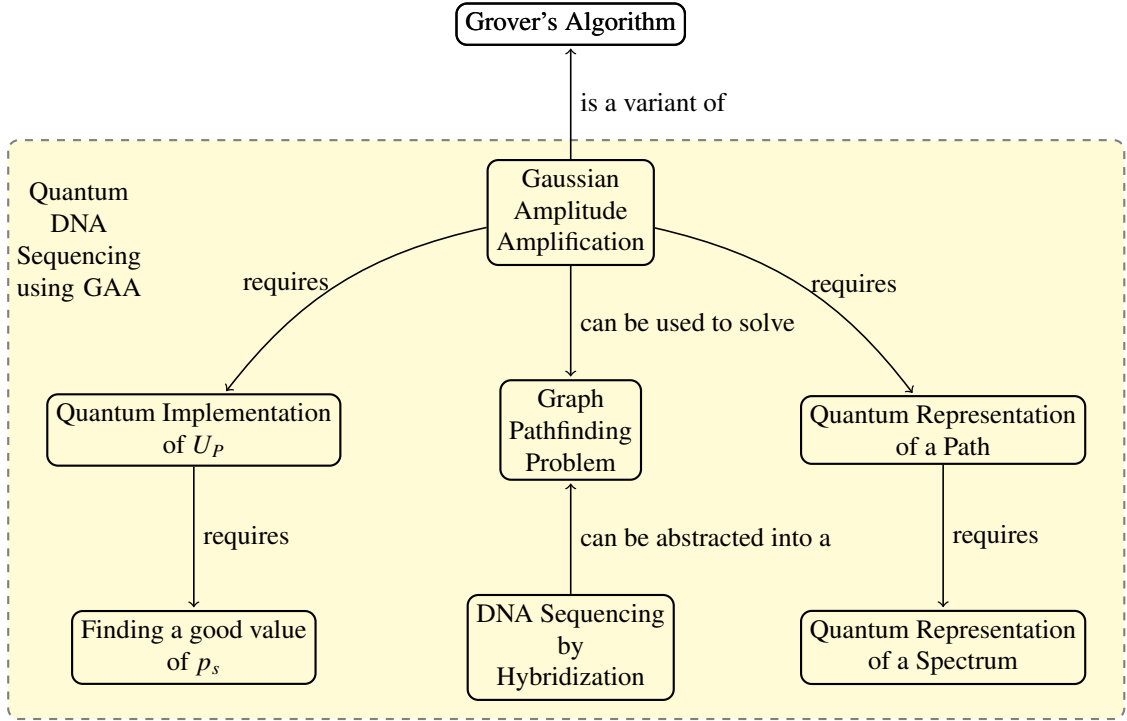
\begin{figure}
	\begin{tikzpicture}[
		thick,
		every matrix/.style={ampersand replacement=\&,column sep=1.5cm,row sep=1.5cm},
		main/.style = {draw, thick, rounded corners, rectangle, align=center},
		to/.style={->,shorten >=1pt, semithick}
		]

		\matrix{
			\&\node[main](4){Grover's Algorithm};\&                                                                                                                           \\
			\&\node[main](1){Gaussian\\ Amplitude\\ Amplification};\&                                                                                                         \\
			\node[main](5){Quantum Implementation\\ of $U_P$};\&\node[main](2){Graph\\ Pathfinding\\ Problem};\& \node[main](7){Quantum Representation\\ of a Path};          \\
			\node[main](6){Finding a good value\\ of $p_s$};\& \node[main](3){DNA Sequencing\\ by\\ Hybridization};\& \node[main](8){Quantum Representation\\ of a Spectrum}; \\
		};
		
		\path (5.west |- 1.north) + (-0.5, 0.25) node (a1) {};
		\path (7.east |- 8.south) + (+0.5, -0.45) node (a2) {};
		\path[fill=yellow!20, rounded corners, draw=black!50, dashed] (a1) rectangle (a2);
		\tikzstyle{texto} = [below, text width=6em, text centered]
		\path (a1.east |- a1.south) + (0.8, -0.3) node (u1)[texto] {Quantum\\ DNA Sequencing\\ using GAA};
		
		\matrix{
			\&\node[main](4){Grover's Algorithm};\&                                                                                                                           \\
			\&\node[main](1){Gaussian\\ Amplitude\\ Amplification};\&                                                                                                         \\
			\node[main](5){Quantum Implementation\\ of $U_P$};\&\node[main](2){Graph\\ Pathfinding\\ Problem};\& \node[main](7){Quantum Representation\\ of a Path};          \\
			\node[main](6){Finding a good value\\ of $p_s$};\& \node[main](3){DNA Sequencing\\ by\\ Hybridization};\& \node[main](8){Quantum Representation\\ of a Spectrum}; \\
		};
		
		\draw[to] (1) -- node[midway, right] {can be used to solve} (2);
		\draw[to] (3) -- node[midway, right] {can be abstracted into a} (2);
		\draw[to] (1) -- node[midway, right] {is a variant of} (4);
		
		\draw[to] (1) to[bend right=20] node[midway, left] {requires} (5);
		\draw[to] (5) -- node[midway, right] {requires} (6);
		
		\draw[to] (1) to[bend right=-20] node[midway, right] {requires} (7);
		\draw[to] (7) -- node[midway, right] {requires} (8);
		
	\end{tikzpicture}
	
	\caption{A concept map of the main ideas that were used to form the methodology that sequences DNA using Gaussian Amplitude Amplification}
	\label{fig:concept_map}
\end{figure}

The main methodology of this study is about turning the DNA Sequencing Problem into a Graph Pathfinding Problem, then using GAA to solve the pathfinding problem which, essentially, solves the DNA Sequencing Problem. A concept map of the main ideas used in this study is shown in Fig. \ref{fig:concept_map}.

\subsection{DNA Sequencing by Hybridization}
\label{sec:SbH}

Sequencing by Hybridization (SbH) involves reconstructing an unknown sequence $N$, with length $n$, from a set of shorter sequences $S$ of length $l$ obtained from a \textit{hybridization experiment}. The set of shorter sequences with length $l$ are called $l$-mers since they are \textit{polymers} of length $l$. In an ideal hybridization experiment, the total number of $l$-mers in the spectrum is $|S|=n-l+1$ \citep{blazewiczDNASequencingPositive1999}. Furthermore, the reconstruction of the sequence $N$ can be done by finding the best arrangement of the $l$-mers contained in the spectrum $S$.

In order to quantify how good a reconstruction is, a connection cost, $c$, and a reconstruction cost $C$ is defined in Eq. (\ref{eq:con_cost}) and Eq. (\ref{eq:path_cost}) respectively. The connection cost quantifies how much information is being sacrificed when connecting two $l$-mers together -- thus, connections with more overlapping strings will have a lower cost. The cost of connecting two $l$-mers, $c$, can be defined as $l$ minus the number of overlapping bases, as shown in Eq. (\ref{eq:con_cost}). Furthermore, the sum of all connections cost, $c$, is the path cost $C$, as shown in Eq. (\ref{eq:path_cost}). By representing each $l$-mer in the spectrum as a node in a graph, sequencing $N$ would be equivalent to finding the Hamiltonian path on the graph with the minimum path cost \citep{blazewiczComplexityDNASequencing2003}.

\begin{equation}
	c = l - (\text{number of overlapping bases})
	\label{eq:con_cost}
\end{equation}

\begin{equation}
	C = \sum_i{c_i}
	\label{eq:path_cost}
\end{equation}

\subsection{Abstracting the DNA Sequencing Problem into a Graph Problem}

The DNA Sequencing problem can be abstracted into a graph problem. Specifically, if the cost functions are defined in Sec. \ref{sec:SbH}, the sequencing problem can be abstracted into finding the Hamiltonian path with the least cost in a graph. Consider the spectrum $S={\text{CT},\text{TT},\text{TG},\text{GA}}$. The spectrum $S$ could be converted into a directional graph as shown in Fig. \ref{fig:planar_graph} by assigning an $l$-mer to each node on the graph and applying the connection costs on the edges.

\subsection{Quantum Representation of a spectrum}

In order to perform quantum computations on the spectrum, a unique state, represented using qubits, must first be assigned to each $l$-mer. Uniquely identifying each $l$-mer in the spectrum can be done by assigning an integer to each $l$-mer. Furthermore, qubits are required to represent integers in a quantum computer; thus, the integers must be in binary form. To calculate the number of qubits required to represent $|S|$ states uniquely, Eq. (\ref{eq:Nq}) is used.

\begin{equation}
	\label{eq:Nq}
	N_q = \lfloor \log_2{(|S|-1)} \rfloor + 1
\end{equation}

Consider the spectrum $S=\{\text{CT}, \text{TT}, \text{TG}, \text{GA}\}$ which contains 4 $l$-mers. Using Eq. (\ref{eq:Nq}), the number of qubits required to label each state in $S$ uniquely is $N_q=2$. Thus, the quantum representation of the spectrum becomes $S=\{\ket{00}, \ket{01}, \ket{10}, \ket{11}\}$ where: $\ket{00} \mapsto \text{CT}$, $\ket{01} \mapsto \text{TT}$, $\ket{10} \mapsto \text{TG}$, and $\ket{11} \mapsto \text{GA}$.

\subsection{Quantum Representation of Paths}
\label{sec:q_representation}

One way of representing a path using qubits is by regarding each node on the graph as a basis state and a sequence of basis states, a path. Consider the sequence $N=\text{CTTGA}$ with the spectrum $S={\text{CT},\text{TT},\text{TG},\text{GA}}$. Representing the elements of the spectrum as nodes in a graph, as shown in Fig. \ref{fig:planar_graph}, we get the path with the least cost: $\text{CT} \to \text{TT} \to \text{TG} \to \text{GA}$. Furthermore, we assign the following basis states for each of the nodes: $\ket{00} \mapsto \text{CT}$, $\ket{01} \mapsto \text{TT}$, $\ket{10} \mapsto \text{TG}$, and $\ket{11} \mapsto \text{GA}$.

\begin{figure}[tb]
	\centering
	\includegraphics[width=15cm]{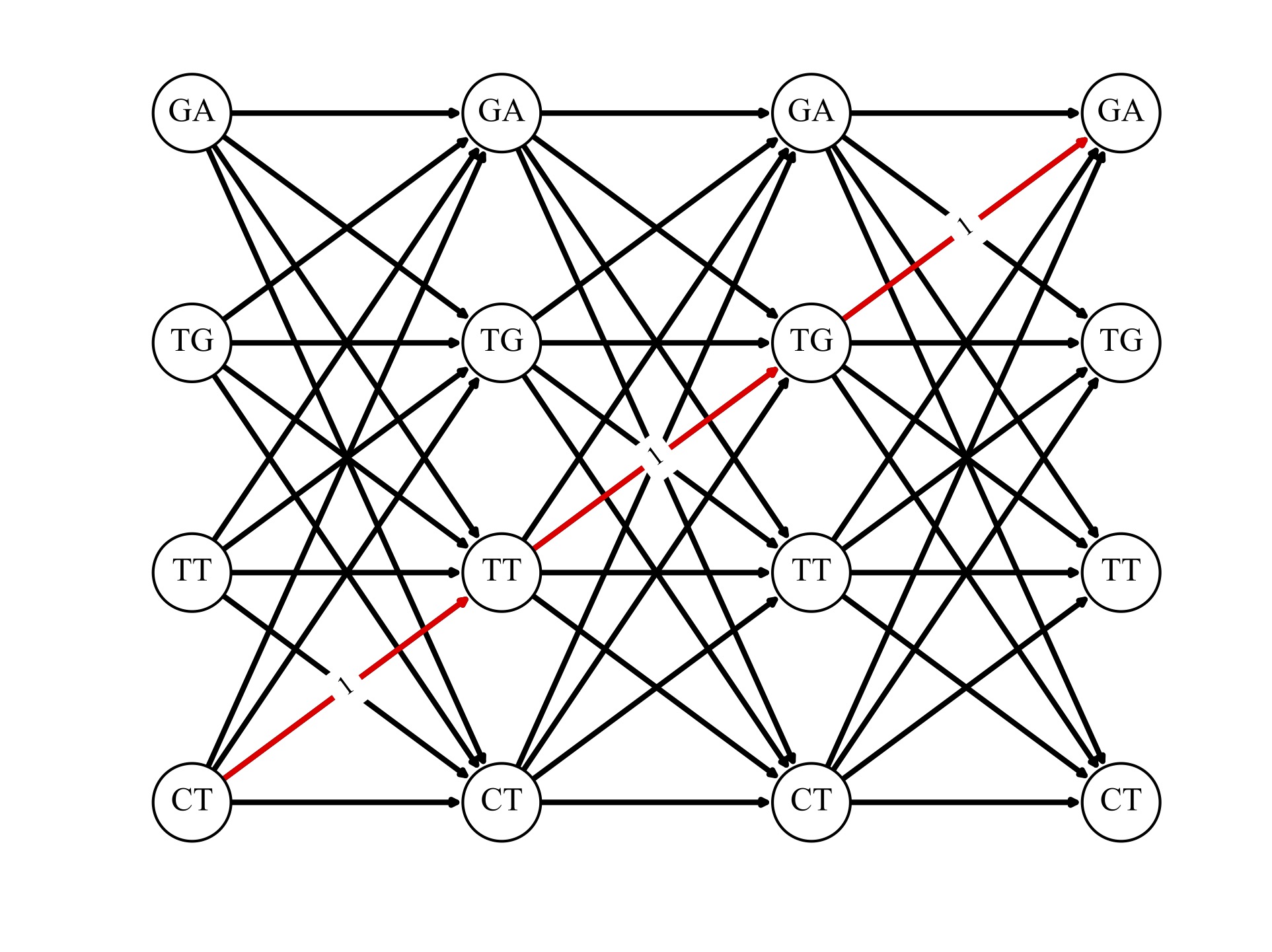}
	\caption{A sequential bipartite graph (SbG) with four layers and four nodes per layer. The path highlighted in red represents the path $\ket{P}=\text{CT} \to \text{TT} \to \text{TG} \to \text{GA}=\ket{00}\bigotimes\ket{01}\bigotimes\ket{10}\bigotimes\ket{11}$.}
	\label{fig:bi_graph}
\end{figure}

To more easily visualize how the path would be represented using qubits, the method that \citet{kochGaussianAmplitudeAmplification2022} used was converting the graph structure into a sequentially connected bipartite graph, as shown in Fig. \ref{fig:bi_graph}. The path highlighted in the figure is $\text{CT}\to\text{TT}\to\text{TG}\to\text{GA}$ which is equivalent to $\ket{P}=\ket{00}\bigotimes\ket{01}\bigotimes\ket{10}\bigotimes\ket{11}$. Note that since every layer will contain $|S|$ nodes, $|S|$ layers, there will be a total of $|P|=|S|^{|S|}$ paths. With this, any path can be represented by a series of basis states. Furthermore, the cost of traversing a path is simply the sum of all the costs between two nodes, as shown in Eq. (\ref{eq:path_cost}). Thus, the cost of the red highlighted path in Fig. \ref{fig:bi_graph} is $C=1+1+1=3$.

\subsection{Grover's Algorithm}

Grover’s algorithm (GA) can be broken down into four steps \citep{kochGaussianAmplitudeAmplification2022, lavorGroverAlgorithmQuantum2003}: 1. The initialization of the $N$ Qubits that will be used: $\ket{\Psi}=\ket{0}^{\bigotimes N}$; 2. The preparation of an equal superposition for all the possible states: $H^{\bigotimes N}\ket{\Psi}=\ket{s}$; 3. The successive application of the oracle ($U_G$) and diffusion ($U_S$) operator for approximately $k\approx \frac{\pi}{4} \sqrt{2^N}$ times; and finally, 4. The measurement of the prepared state. The general idea of how GA works is by using an oracle to apply a phase to the marked state/s and then applying the diffusion operator to reflect the state about the average amplitude without collapsing the prepared superposition. The measurement of the prepared superposition collapses it. Furthermore, it is expected with a high probability to measure a marked state after the successive application of $U_G$ and $U_S$, $k$-times. Shown in Eq. (\ref{eq:U_G}) is the Grover Oracle and in Eq. (\ref{eq:U_S}), the Diffusion operator. The pseudocode for Grover’s algorithm is also listed in Alg. \ref{alg:grover}.

\begin{equation}
	U_G = \begin{cases}
		e^{i\pi}\ket{\Psi_i} & \text{marked state}   \\
		\ket{\Psi_i}         & \text{unmarked state}
	\end{cases}
	\label{eq:U_G}
\end{equation}

\begin{equation}
	U_S = 2\ket{s}\bra{s} - \mathbb{I}
	\label{eq:U_S}
\end{equation}

\begin{algorithm}
	\caption{Grover's Algorithm}
	\label{alg:grover}
	
	\begin{algorithmic}[1]
		\STATE Initialize Qubits: $\ket{\Psi}=\ket{0}^{\bigotimes N}$
		\STATE Prepare Equal Superposition: $H^{\bigotimes N}\ket{\Psi}=\ket{s}$
		\FOR{$k=\bigg\lfloor\frac{\pi}{4}\sqrt{2^N}\bigg\rfloor$}
		\STATE Apply $U_G\ket{\Psi}$ (Oracle)
		\STATE Apply $U_S\ket{\Psi}$ (Diffusion)
		\ENDFOR
		\STATE Measure $\ket{\Psi}$
	\end{algorithmic}
\end{algorithm}

\subsection{Gaussian Amplitude Amplification for Pathfinding}
\label{sec:GAA}

The steps for Gaussian Amplitude Amplification are almost precisely the same as Grover’s algorithm except for using a different oracle, $U_P$, shown in Alg. \ref{alg:GAA} \citep{kochGaussianAmplitudeAmplification2022}. $U_P$ is a diagonal matrix as shown in Eq. (\ref{eq:U_P}). The purpose of $U_P$ is to apply a phase to each path in the prepared superposition proportional to their costs multiplied by some constant, $p_s$. The constant $p_s$ is used to scale the phases into the range $[x,x+2\pi]$ and will be discussed further in Sec. \ref{sec:ps} and \ref{sec:ps_optimization}. The number of iterations, $k$, is adapted from Grover's Algorithm, where $k$ is proportional to the square root of the total number of states. \citet{boyerTightBoundsQuantum1998} suggests that for cases with $N$ total states and $t$ optimal solutions, the value of $k$ is bounded as shown in Eq. (\ref{eq:k_1}) where $m$ is some unknown integer. However, since in the case of this study, it is frequently the case that $t \ll N$; thus, Eq. (\ref{eq:k_1}) can be simplified int. Eq. (\ref{eq:k_2}). Thus, the optimal value of $k$ becomes Eq. (\ref{eq:k_opt}).

\begin{algorithm}
	\caption{Gaussian Amplitude Amplification}
	\begin{algorithmic}[1]
		\STATE Initialize Qubits: $\ket{\Psi}=\ket{0}^{\bigotimes N_Q}$
		\STATE Prepare Equal Superposition: $H^{\bigotimes N_Q}\ket{\Psi}=\ket{s}$
		\FOR{$k=\bigg\lfloor\frac{\pi}{4}\sqrt{{|S|}^{|S|}}\bigg\rfloor$}
		\STATE Apply $U_P\ket{\Psi}$ (Oracle)
		\STATE Apply $U_S\ket{\Psi}$ (Diffusion)
		\ENDFOR
		\STATE Measure $\ket{\Psi}$
	\end{algorithmic}
	\label{alg:GAA}
\end{algorithm}

\begin{equation}
	U_P =
	\begin{pmatrix}
		e^{i p_s\cdot C_0} & \hdots & 0                       \\
		\vdots             & \ddots & \vdots                  \\
		0                  & \hdots & e^{i p_s \cdot C_{|P|}}
	\end{pmatrix}_s
	\label{eq:U_P}
\end{equation}

\begin{equation}
	m \le k \le \frac{\pi}{4} \sqrt{\frac{N}{t}}
	\label{eq:k_1}
\end{equation}

\begin{equation}
	m \le k \le \frac{\pi}{4} \sqrt{N}
	\label{eq:k_2}
\end{equation}

\begin{equation}
	k = \bigg\lfloor \frac{\pi}{4} \sqrt{N} \bigg\rfloor = \bigg\lfloor \frac{\pi}{4} \sqrt{{|S|}^{|S|}} \bigg\rfloor
	\label{eq:k_opt}
\end{equation}

\subsection{Quantum Circuit Implementation of $U_P$}

While applying $U_P$ is the desired effect for the oracle, it cannot be constructed without calculating the costs of the paths beforehand –- which is expensive computationally and the step we wish to skip through utilizing the advantage of using quantum computers. Implementing Up in an actual quantum circuit requires a smaller operator $U_{ij}$, shown in Eq. (\ref{eq:U_{ij}}), which encodes all of the phases between consecutive layers $i$ and $j$ (from the sequential bipartite graph) is introduced \citep{kochGaussianAmplitudeAmplification2022}. The phases applied in the operator $U_{ij}$ is equal to the cost of connecting a node, $\mu$, in layer $i$ to a node, $\nu$, in layer $j$ multiplied by some constant $p$ as shown in Eq. (\ref{eq:phi_uv}).

\begin{equation}
	\label{eq:U_{ij}}
	U_{ij}=
	\begin{pmatrix}
		e^{i\phi_{00}} & 0              & 0              & 0              \\
		0              & e^{i\phi_{01}} & 0              & 0              \\
		0              & 0              & e^{i\phi_{10}} & 0              \\
		0              & 0              & 0              & e^{i\phi_{11}}
	\end{pmatrix}
\end{equation}

\begin{equation}
	\label{eq:phi_uv}
	\phi_{\mu\nu} = p_s \cdot c_{\mu\nu}
\end{equation}

As of writing this study, choosing the constant $p$ is still an ongoing area of research, and no reliable method exists for finding an optimal $p$ since it seems to vary depending on the structure of the graph. Thus, a classical optimization method will be used to find an optimal $p$. More on this on Sec. \ref{sec:ps_optimization}.

\subsection{DNA Sequencing using Gaussian Amplitude Amplification}
Using Gaussian Amplitude Amplification (GAA) for DNA sequencing is done by reframing the sequencing problem into a pathfinding problem and applying GAA. Consider the sequence $N=\text{GGATG}$ with the spectrum $S=\{\text{GG},\text{GA},\text{AT},\text{TG}\}$. A graph could be constructed using the given spectrum and is shown in Fig. \ref{fig:planar_graph}. After converting the graph into sequential bipartite graphs, as described in Sec. \ref{sec:q_representation} (and shown in Fig. \ref{fig:bi_graph}), applying GAA (as prescribed in Sec. \ref{sec:GAA}) becomes a straightforward task.

\subsection{The scaling factor $p_s$}
\label{sec:ps}

The oracle $U_P$ applies a unique phase on all of the paths in the superposition state. Each phase applied to a path is proportional to its path cost, $C$. In order for GAA's amplification process to work optimally, all phases of the states within the prepared superposition must be within the range $[x, x+2\pi]$. One way of constraining all the phases within the specified range is by multiplying all the phases by some constant $p_s$. This method, however, requires finding a good value of $p_s$; thus, finding a good value for $p_s$ shall be discussed in Sec. \ref{sec:ps_optimization}.

\subsection{Finding a good value of $p_s$}
\label{sec:ps_optimization}

Currently, there is no straightforward way of finding $p_s$. It also does not help that $p_s$ will vary depending on the given graph. \citet{kochGaussianAmplitudeAmplification2022} suggests that Eq. (\ref{eq:init_ps}) could be used as an initial value for $p_s$. However, since the phase is periodic, good values for $p_s$ can be found by simply checking values between $0$ and $2\pi$. To quantify how good a value of $p_s$ is, a \textit{cost}, $\delta$, is defined. Equation (\ref{eq:ps_cost}) shows how $\delta$ is calculated. $\delta$ is defined to be the percentage of the measured costs that \textbf{is not} the minimal cost. Thus a value of $p_s$ that minimizes $\delta$ maximizes the percentage of the measured costs that \textbf{is} the minimal cost.

\begin{equation}
	\label{eq:init_ps}
	p_s = \frac{2\pi}{\text{max path weight} - \text{min path weight}}
\end{equation}

\begin{equation}
	\label{eq:ps_cost}
	\delta = \bigg(1-\frac{\text{number of paths measured with minimal cost}}{\text{total number of paths measured}}\bigg) \times 100
\end{equation}

\section{Methodology}
\label{sec:methodlogy}
The reconstruction of a DNA sequence using Sequencing by Hybridization (SbH) is equivalent to finding the Hamiltonian path in a graph.  A quantum pathfinding algorithm called the Gaussian Amplification Algorithm (GAA) \citep{kochGaussianAmplitudeAmplification2022} will be used to solve the computational part of SbH.  Chapter \ref{ch:gen_spec} will outline the steps in generating the spectrum for an ideal hybridization experiment.  Chapter \ref{ch:gen_diGraph}  will discuss how a directed graph shall be constructed using the generated spectrum.  Since GAA, as \citet{kochGaussianAmplitudeAmplification2022} described in their paper, requires a particular graph structure. Chapter \ref{ch:gen_biGraph} will discuss converting the directed graph in Ch. \ref{ch:gen_diGraph} into the required graph structure.  Finally, chapter \ref{ch:qcircuit} will discuss how the quantum circuit used for GAA will be constructed.

\subsection{Generating the Dataset}
\label{ch:gen_spec}

To generate a DNA strand, $N$, of length $n$, a base from the four nucleobases is chosen at random $n$ times.  Thus, for a strand of length $n$, the total possible number of strands that could be generated is $4^l$.  To generate the spectrum, $S$, from $N$ –- slices of length $l$ are cut from the generated $N$.  Note that only an ideal hybridization experiment will be considered for this study –-  thus, the total number of $l$-mers must be $|S|=n-l+1$.  Consider the randomly generated sequence $N=\text{GGATG}$ where $n=5$.  Setting $l=2$ should yield the spectrum $S={\text{GG},\text{GA},\text{AT},\text{TG}}$, which has four $l$-mers.

\subsection{Constructing A Directed Graph From the Spectrum}
\label{ch:gen_diGraph}

The graph for a given spectrum, $S$, uses the elements of $S$ as the nodes of the graph.  Furthermore, the cost of the graph, $C$, is the length of an $l$-mer, $l$, minus the number of bases that overlap when two $l$-mers are joined (see Eq. (\ref{eq:con_cost})).

Note that the order in joining two $l$-mers is directional, e.g., joining $\text{GG}$ to $\text{GA}$ has a cost of $1$ while joining $\text{GA}$ to $\text{GG}$ has a cost of $2$.  Using the rules discussed, a graph using the spectrum $S=\{\text{GG},\text{GA},\text{AT},\text{TG}\}$ is constructed as shown in Fig. \ref{fig:planar_graph}.  Following the path with the least cost (highlighted in red), the sequence $N=\text{GGATG}$ can be reconstructed. 

\subsection{Converting the Directed Graph Into a Sequential Bipartite Graph}
\label{ch:gen_biGraph}

To solve the DNA sequencing problem using Gaussian Amplitude Amplification, the graph from Ch. \ref{ch:gen_diGraph} must first be converted into a sequential bipartite graph.  The first step involves assigning a basis state for each of the $l$-mers in the spectrum.  Given the spectrum $S=\{\text{GG},\text{GA},\text{AT},\text{TG}\}$, the following assignments can be performed: $\text{GG}\mapsto\ket{00}$, $\text{GA}\mapsto\ket{01}$, $\text{AT}\mapsto\ket{10}$, $\text{TG}\mapsto\ket{11}$.  Furthermore, a spectrum with $|S|$ elements will always require a sequential bipartite graph with $|S|$ layers and $|S|$ nodes per layer, as shown in Fig. \ref{fig:bi_graph}.  Since the graph shown in Fig. \ref{fig:bi_graph} has connections of $l$-mers with itself, the cost function in Eq. (\ref{eq:con_cost}) will no longer be applicable since it would assign a cost of $0$ for paths that lead to the same node.  Thus, we use Eq. (\ref{eq:con_cost2}), which assigns a cost $l$ for paths that connect the same node.  The basis states assigned to each node will be used to specify a path in the graph; for example, the path $\text{GG}\to\text{GA}\to\text{AT}\to\text{TG}$ will be encoded as $\ket{P}=\ket{00}\bigotimes\ket{01}\bigotimes\ket{10}\bigotimes\ket{11}=\ket{00011011}$.

\begin{equation}
	\label{eq:con_cost2}
	C = \begin{cases}
		l - \text{overlap} & \text{if }l\text{-mer A} \not\equiv l\text{-mer B} \\
		l & \text{if }l\text{-mer A} \equiv l\text{-mer B}
	\end{cases}
\end{equation}

\subsection{Quantum Circuit}
\label{ch:qcircuit}

\begin{figure}[tb]
	\centering
	\includegraphics[width=14cm]{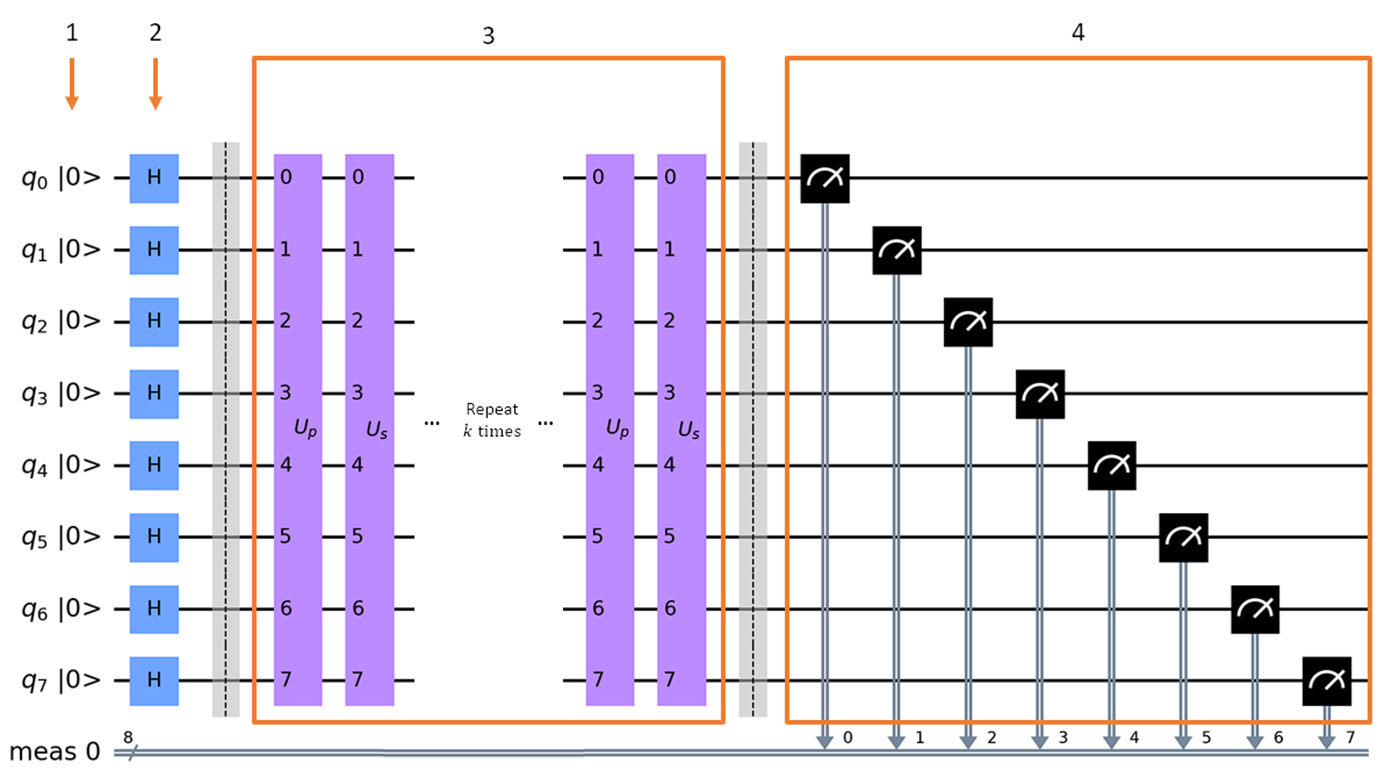}
	\caption{An illustration of the quantum circuit that will be used to run GAA.  The first layer corresponds to the qubit initialization, the second is the preparation of the equal superposition, the third is the sequential application of $U_P$ and $U_S$, and finally, the measurement is conducted in the fourth layer.}
	\label{fig:qcirc}
\end{figure}

In this section, we consider building the circuit required to use Gaussian Amplitude Algorithm (GAA) to solve the graph in Fig. \ref{fig:bi_graph}, which has four layers and four nodes per layer.  GAA is composed of 4 steps: 1.   Initialization of the Qubits; 2.   Preparation of the Superposition; 3.  Successive application of the Oracle and Diffusion operator; and finally, 4.  Measurement of the results. See Fig. \ref{fig:qcirc} for an illustration of the quantum circuit.

\subsection{Initialization of the Qubits}

The initialization of qubits is the preparation of the state $\ket{\Psi}$ with all $N_Q$ qubits set to $\ket{0}$ ($\ket{\Psi}=\ket{0}^{\bigotimes N_Q}$).  Calculating $N_Q$  requires knowing the number of qubits required to describe all $l$-mers uniquely ($N_q$) and the number of layers in the graph, $L$.  $N_q$ can be calculated using Eq. (\ref{eq:Nq}).  In the case of $|S|=4$: $N_q=\lfloor \log_2{4-1} \rfloor + 1=2$ qubits are required to describe each $l$-mer as a basis state.  Furthermore, the number of layers in the graph is $L=|S|$.  Thus, there will be a total of $N_Q$ qubits (See Eq. (\ref{eq:NQ})).  In the case of $|S|=4$: $N_Q=2\times 4=8$ qubits are required to run GAA.  Thus, for $N_Q=8$: $\ket{\Psi}=\ket{0}^{\bigotimes 8}$.

\begin{equation}
	\label{eq:NQ}
	N_Q = N_q \times L
\end{equation}

\subsection{Preparation of the Superposition}

The preparation of the equal superposition states, $\ket{s}$, will be done by applying Hadamard Gates on all the initialized qubits.  A Hadamard gate can be written in matrix form, as shown in Eq. \ref{eq:H_gate}.  Thus, to create the superposition state: $\ket{s}=H^{\bigotimes N_Q}\ket{\Psi}$.

\begin{equation}
	\label{eq:H_gate}
	H =
	\frac{1}{\sqrt{2}}
	\begin{pmatrix}
		1 & 1 \\
		1 & -1
	\end{pmatrix}
\end{equation}

\subsection{Phase Oracle}

To construct the phase oracle, $U_P$, applies a phase on all the possible states on the superposition $\ket{s}$.  Since $U_P$ cannot be constructed without calculating the cost of a whole path $\ket{P}$, we introduce a smaller operator $U_{ij}$ which applies a phase on the states in layers $i$ and $j$.  The successive applications of $U_{ij}$ builds up to $U_P$ (Eq. \ref{eq:U_P2}).

\begin{equation}
	\label{eq:U_P2}
	U_P = \prod_{i=1}^{L-1}{U_{i,i+1}}
\end{equation}

\subsection{Diffusion Operator}

The diffusion operator reflects every state about the average amplitude without computing the average itself –- it is what makes it possible to boost the amplitudes of the desired states.  The diffusion operator is defined as shown in  Eq. (\ref{eq:U_S}).

\subsection{Measurement}

The measurement refers to performing a physical measurement of the prepared state $\ket{s}$.  Assuming that the previous steps were performed and a good value of $p$ was used to construct $U_{ij}$, the state that would be measured should have a high probability of being the optimal path $\ket{P_\text{opt}}$.

\subsection{Finding an optimal $p_s$}

The constant $p_s$ is a scaling factor for the phases applied by the oracle, $U_P$.  Since the phase could only be periodic between $0$ and $2\pi$, a good value for $p_s$ could be found by simply checking values within that range. To quantify how good $p_s$ is, Eq. (\ref{eq:ps_cost}) is used.  Equation (\ref{eq:ps_cost}) calculates the percentage of the measured costs that is not the minimal cost.  A good value for $p_s$ could be found by minimizing Eq. (\ref{eq:ps_cost}).

\section{Results and Discussion}
\label{sec:results}

\begin{figure}[h]
	\centering
	\includegraphics[width=12cm]{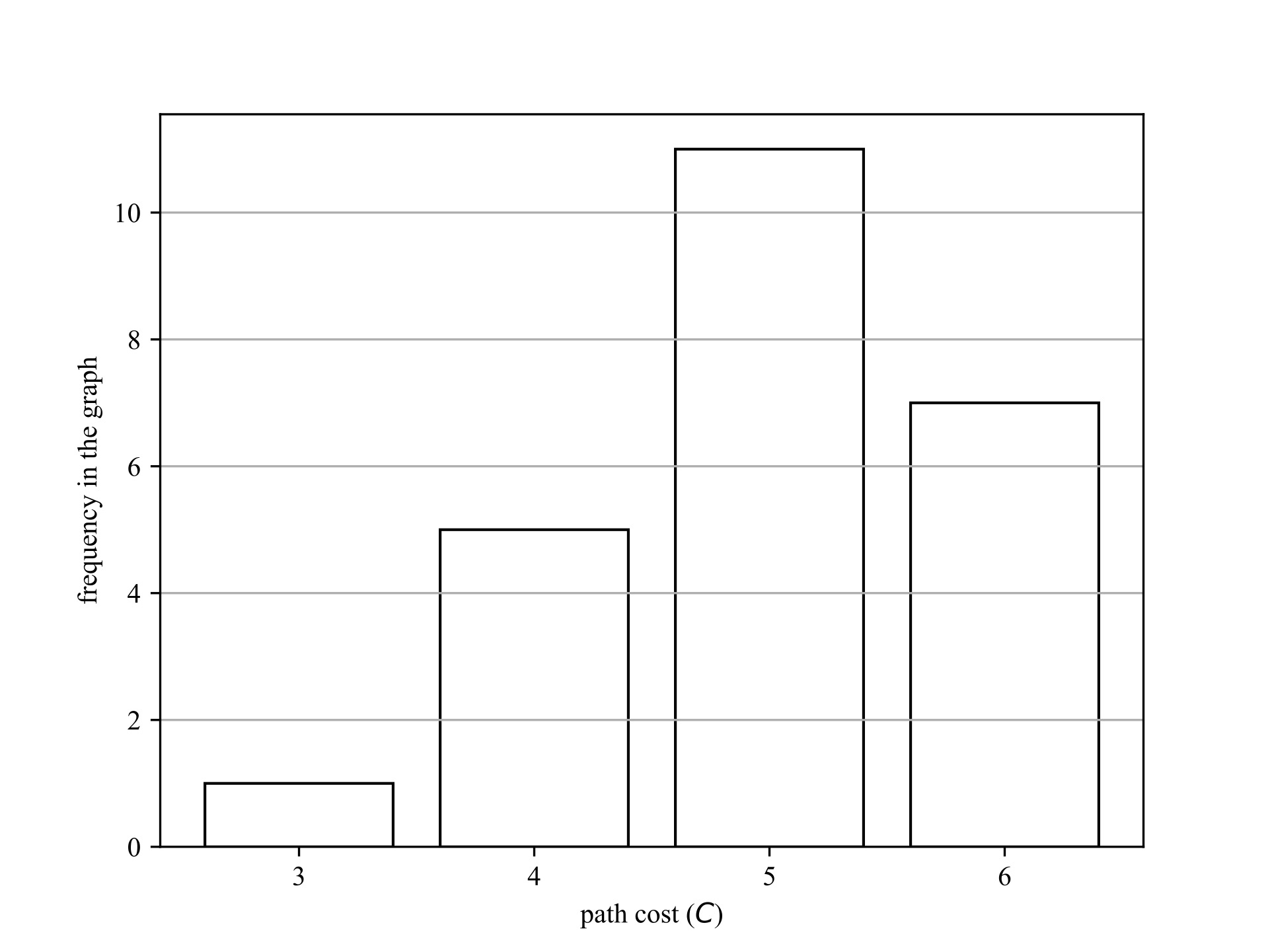}
	\caption{A histogram of the costs of all the paths in the generated graph.}
	\label{fig:hist1}
\end{figure}

For the results of our simulation, we consider the generated DNA strand, $N=\text{CTTGA}$, where the spectrum is given by $S=\{\text{CT},\text{TT},\text{TG},\text{GA}\}$ for $l=2$. The corresponding graph that is built using this spectrum was shown earlier in Fig. \ref{fig:planar_graph}. On the other hand, the equivalent sequential bipartite graph looks similarly as Fig.~\ref{fig:bi_graph}. Now in Table \ref{tab:freq_tables} we provide  a frequency table of the path costs, $C$, obtained from the SbG, and the associated histogram is shown in Fig. \ref{fig:hist1}. From the frequency table, it is evident that if a path is chosen randomly, the probability of selecting a path with the least cost is $p=1/24$, i.e. there is only one path out of 24 paths with the least cost of 3.

\begin{table}[h]
	\centering
	\begin{tabular}{cccc}
		\hline
		& frequency & \multicolumn{2}{c}{mean frequency}              \\
		Path Cost ($C$) & All Paths & $p_s=2.09$                         & $p_s=4.88$ \\
		\hline
		3               & 1         & 3.99                               & 709.17     \\
		4               & 5         & 144.25                             & 134.77     \\
		5               & 11        & 449.15                             & 74.13      \\
		6               & 7         & 402.61                             & 81.94      \\
		\hline
		Total           & 24        & 1000                               & 1000.01    \\
		$\delta(\%)$        & 95.83     & 99.60                              & 29.08      \\
		\hline
	\end{tabular}
	\caption{A frequency table for the paths with a certain path cost, $C$. The second column is the frequency from simply counting all the possible paths. On the other hand, the third and fourth columns are the mean frequencies from 1000 simulations with 1024 shots each.}
	\label{tab:freq_tables}
\end{table}

\begin{figure}[h!]
	\centering
	\includegraphics[width=12cm]{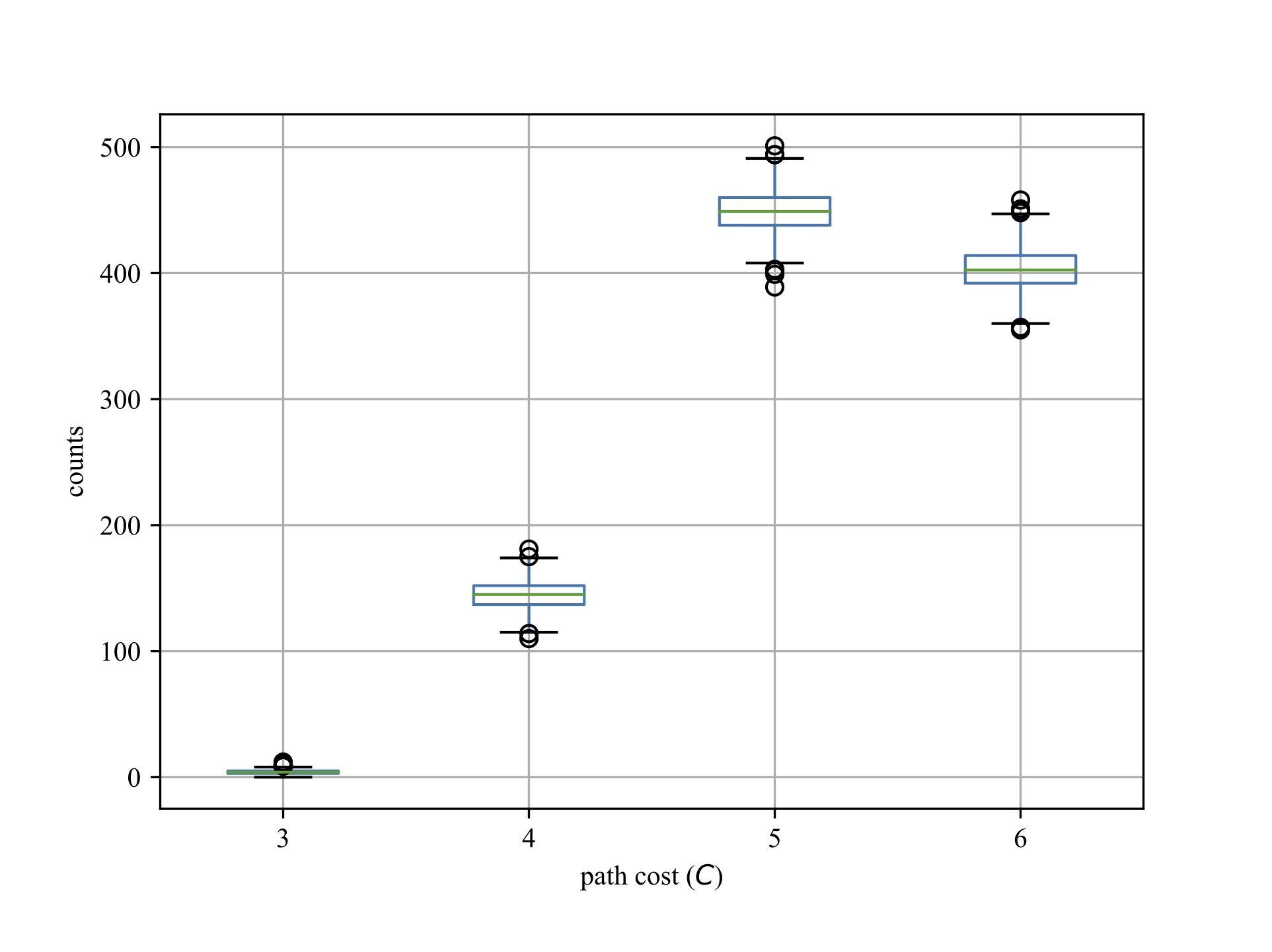}
	\caption{A box plot of the costs associated with the measured paths after 1000 trials with $p_s=2.09$. Each trial is a simulation applying GAA with 1024 shots. The mean counts of each path costs: 3, 4, 5, and 6 are 3.99, 144.25, 449.15, and 402.61 respectively. Additionally, their standard deviations are 1.97, 11.11, 16.50, and 16.28 respectively. The data from this plot suggests that it is likely that the path cost of a path yielded from a simulation after applying GAA with $p_s=2.09$ is $5$.}
	\label{fig:hist2}
\end{figure}

\begin{figure}[h!]
	\centering
	\includegraphics[width=11.5cm]{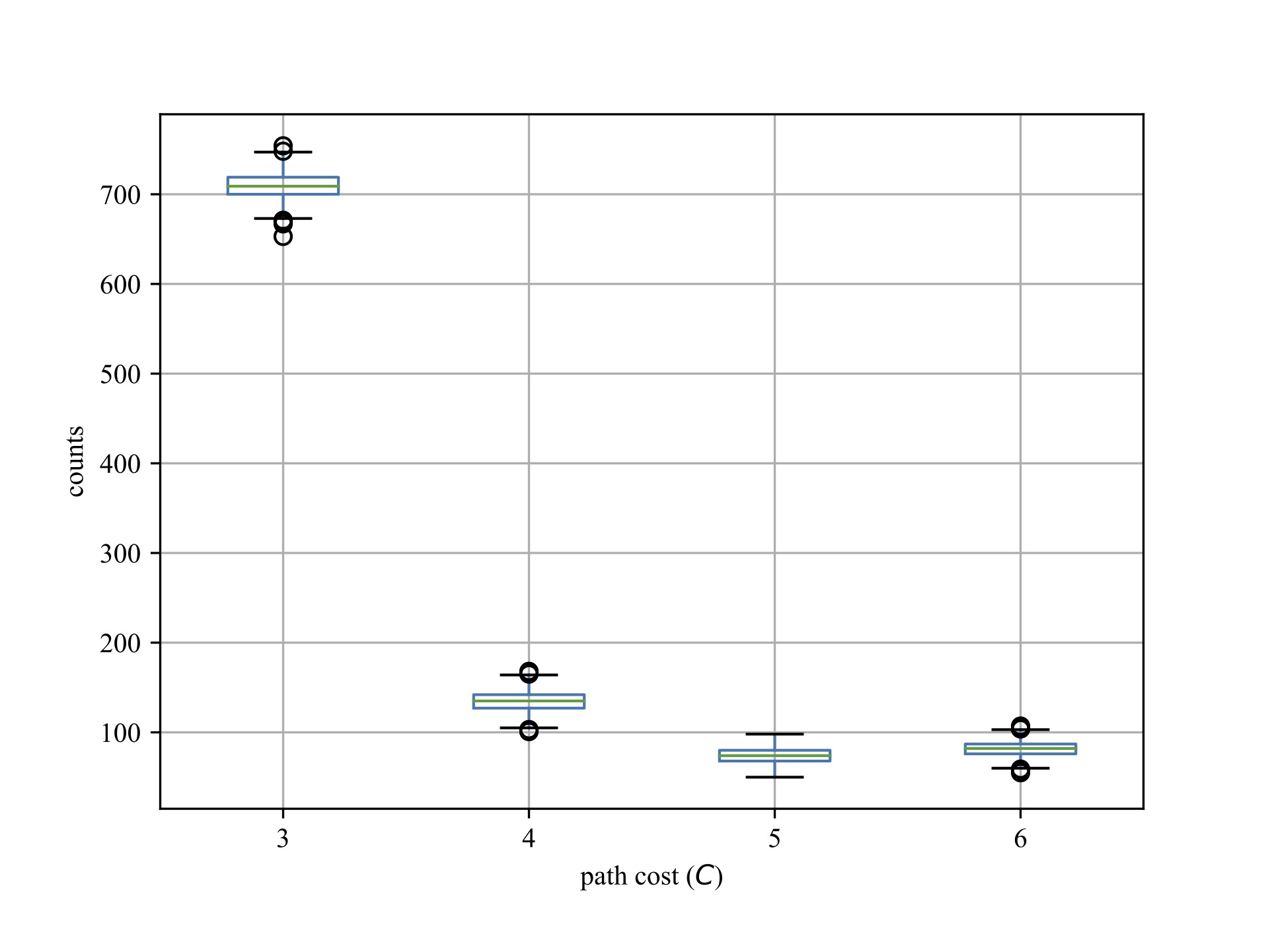}
	\caption{A box plot of the costs associated with the measured paths after 1000 trials with $p_s=4.88$. Each trial is a simulation applying GAA with 1024 shots. The mean counts of each path costs: 3, 4, 5, and 6 are 709.17, 134.77, 74.13, and 81.94 respectively. Additionally, their standard deviations are 13.92, 10.81, 8.17, and 8.53 respectively. The data from this plot suggests that it is likely that the path cost of a path yielded from a simulation after applying GAA with $p_s=4.88$ is $3$.}
	\label{fig:hist3}
\end{figure}

\begin{figure}[h!]
	\centering
	\includegraphics[width=11.5cm]{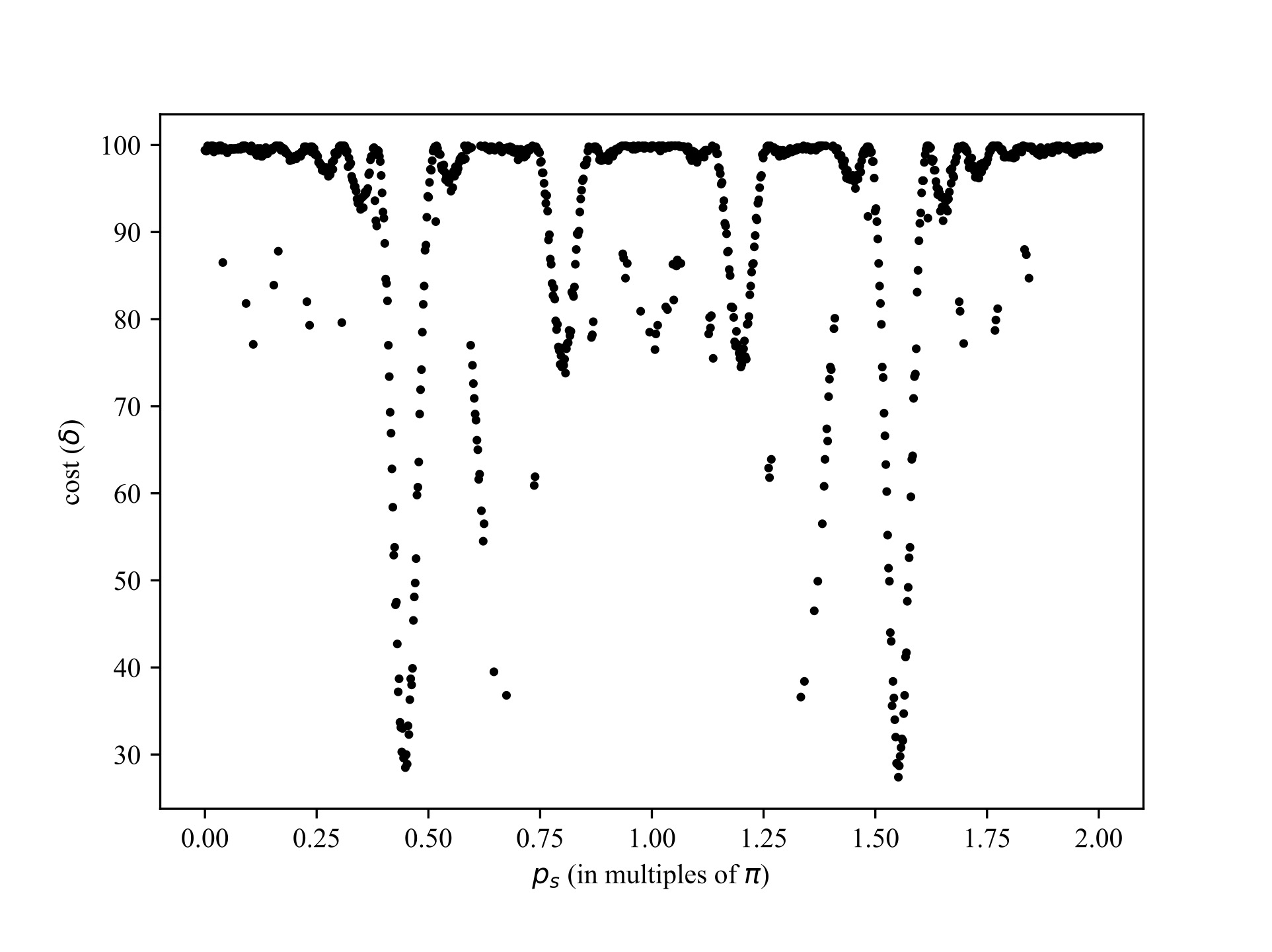}
	\caption{A plot showing values of $p_s$ that were tested and their associated costs, $\delta$}.
	\label{fig:find_ps}
\end{figure}

Applying the GAA with a $p_s$ calculated using Eq.~(\ref{eq:init_ps}), a value of $p_s=2.09$ is found. However, as shown in Table \ref{tab:freq_tables} and in Fig. \ref{fig:hist2}, we notice that the probability that gets  largely boosted is the path with a cost of 5, instead of 3. With this result, we realize that there is a need to optimize $p_s$. In order to do this, the cost $\delta$ is minimized by testing values within the range of $[0, 2\pi]$ as shown in Fig. \ref{fig:find_ps}. It was found that $p_s=4.88$ gives significantly better results. As presented in Table \ref{tab:freq_tables} and in Fig. \ref{fig:hist3}, with $p_s = 4.88$ amplifies the probability of the path with the least cost of 3. Furthermore, the $\delta$ for using $p_s=4.88$ is $29.08$ which implies that a path, $\ket{P}$, with minimal cost will be found approximately $1- 29.08\% = 70.92\%$ of the time after GAA is applied.

To sum up, we have shown that if a path is chosen randomly (with $p = 1/24 = 0.0417$), it is expected to measure a path having a minimal cost after 24 queries on average. On the other hand, after applying the method discussed in this study (with $p = 0.7092$), we can anticpate to measure a path with  a least cost after 1.41 queries on average. Hence, the use of GAA has significantly lessened the required number of queries by $(24-1.41)/24 = 94.13\%$.

\section{Conclusions}
\label{sec:conclusions}

The Genome Sequencing Problem can be abstracted into a graph problem of finding the Hamiltonian Path with the least cost. In order to do this, a directed graph was formed by setting each node of the graph to be an $l$-mer from the spectrum and the edge weights being the cost of connecting two $l$-mers. The constructed directed graph was then converted into sequential bipartite graphs (SbG) to prepare it for the Gaussian Amplitude Amplification (GAA). After testing 1000 values of $p_s$ between $[0, 2\pi]$, it was found it is when $p_s = 4.88$ that amplifies the probability of measuring a path with the least cost at 70.92\% for the case when $l = 2$ and spectrum size of $|S| = 4$. This result shows a 94.21\% improvement in the required number of queries compared to randomly choosing a path . Thus, this suggests that GAA could be a viable method to be used for genome sequencing.

This study has yielded promising results for the ideal case of SbG where the experimental stage was able to gather all the possible $l$-mers of length $l$. However, in practice, it is possible for the experimental stage to not detect an $l$-mer or yield an $l$-mer which should not actually be in the spectrum. These kinds of errors are called negative and positive errors respectively. Additionally, the length of the DNA strand sequenced in this study was limited to 5 and the $l$-mer length to 2. Hence, a suggestion for future studies is to include both the positive and negative errors into account, as well as considering longer DNA strands or $l$-mers.

\vspace{1cm}
\noindent {\bf Acknowledgement}\\
\\
\indent This research work and the author Dr. Carlos Baldo III were supported by the Associate Fellow Program of the APCTP through the Science and Technology Promotion Fund and Lottery Fund of the Korean Government. This work also received an additional support from the Korean Local Governments— Gyeongsangbuk-do Province and Pohang City.

\FloatBarrier
\bibliographystyle{elsarticle-num-names}
\bibliography{bibfile}

\end{document}